\documentclass[sigconf]{acmart}
\usepackage{multirow} 
\usepackage{amsmath}  
\AtBeginDocument{%
  }

\setcopyright{acmlicensed}
\copyrightyear{2018}
\acmYear{2018}
\acmDOI{XXXXXXX.XXXXXXX}
\acmConference[KDD '26]{The 32nd ACM SIGKDD Conference on Knowledge Discovery and Data Mining}{August, 2026}{TBA}
\acmISBN{978-1-4503-XXXX-X/2018/06}

\begin{document}

\title{DTN:\,\,Deep Multiple Task-specific Feature Interactions Network for Multi-Task Recommendation}

\author{Yaowen Bi}
\authornote{Both authors contributed equally to this research. Corresponding author is Yaowen Bi.}

\author{Yuteng Lian}
\authornotemark[1]
\email{yaowen.bi@shopee.com}
\email{yuteng.lian@shopee.com}
\affiliation{%
  \institution{Shopee Pte. Ltd., Singapore}
  \country{Singapore}
}

\author{Jie Cui}
\authornote{This author is the one who did the really hard work for Online Testing and Visualization Analysis.}
\author{Jun Liu}
\email{jie.cui@shopee.com}
\email{jun.liu@shopee.com}
\affiliation{%
  \institution{Shopee Pte. Ltd., Singapore}
  \country{Singapore}
}

\author{Peijian Wang}
\author{Guanghui Li}
\email{peijian.wang@shopee.com}
\email{guanghui.li@shopee.com}
\affiliation{%
  \institution{Shopee Pte. Ltd., Singapore}
  \country{Singapore}
}

\author{Xuejun Chen}
\author{Jinglin Zhao}
\author{Hao Wen}
\author{Jing Zhang}
\email{xuejun.chen@shopee.com}
\email{jinglin.zhao@shopee.com}
\email{hao.wen@shopee.com}
\email{jing.zhang@shopee.com}
\affiliation{%
  \institution{Shopee Pte. Ltd., Singapore}
  \country{Singapore}
}

\author{Zhaoqi Zhang}
\author{Wenzhuo Song}
\author{Yang Sun}
\author{Weiwei Zhang}
\email{zhaoqi.zhang@shopee.com}
\email{wenzhuo.song@shopee.com}
\email{yang.sun@shopee.com}
\email{zhangweiwei@shopee.com}
\affiliation{%
  \institution{Shopee Pte. Ltd., Singapore}
  \country{Singapore}
}

\author{Mingchen Cai}
\author{Jian Dong}
\author{Guanxing Zhang}
\email{mark.cai@shopee.com}
\email{jian.dong@shopee.com}
\email{andy.zhanggx@shopee.com}
\affiliation{%
  \institution{Shopee Pte. Ltd., Singapore}
  \country{Singapore}
}

\renewcommand{\shortauthors}{Bi et al.}

\begin{abstract}
Neural-based multi-task learning (MTL) has been successfully applied to many recommendation applications. However, these MTL models (e.g., MMoE, PLE) did not consider feature interaction during the optimization, which is crucial for capturing complex high-order features and has been widely used in ranking models for real-world recommender systems. Moreover, through feature importance analysis across various tasks in MTL, we have observed an interesting divergence phenomenon that the same feature can have significantly different importance across different tasks in MTL. To address these issues, we propose Deep Multiple Task-specific Feature Interactions Network (DTN) with a novel model structure design. DTN introduces multiple diversified task-specific feature interaction methods and task-sensitive network in MTL networks, enabling the model to learn task-specific diversified feature interaction representations, which improves the efficiency of joint representation learning in a general setup. We applied DTN to Shopee's real-world E-commerce recommendation dataset, which consisted of over 6.3 billion samples, the results demonstrated that DTN significantly outperformed state-of-the-art MTL models. Moreover, during online evaluation of DTN in a large-scale E-commerce recommender system, we observed a 3.28\% in clicks, a 3.10\% increase in orders and a 2.70\% increase in GMV (Gross Merchandise Value) compared to the state-of-the-art MTL models. Finally, extensive offline experiments conducted on public benchmark datasets demonstrate that DTN can be applied to various scenarios beyond recommendations, enhancing the performance of ranking models.
\end{abstract}

\ccsdesc[500]{Information systems~Information retrieval}

\keywords{Recommender System, Multi-task Learning, Feature Interaction}


\maketitle

\section{Introduction}

Recommender systems (RS) plays a crucial role in user applications by providing personalized recommendations for preferred candidates such as items, ads, videos, news, and more, based on user properties. In an industrial setting, RS needs to fulfill various objectives, such as Click-Through Rate (CTR), Add-To-Cart rate (ATC), Conversion Rate (CVR), and so on. Multi-task learning (MTL), which involves training a single network to generate multiple predictions for different objectives, is well-suited to meet these requirements. Consequently, there is an increasing trend in applying MTL to recommender systems (RS) to simultaneously model multiple aspects, including CTR, CVR and other objectives. In fact, this approach has become the predominant method in major industry applications \cite{ma2018entire,caruana1997multitask,ma2018modeling,maziarz2019gumbel,tang2020progressive}.

MTL has been demonstrated to enhance learning efficiency by facilitating information sharing among the tasks \cite{caruana1997multitask}. However, these MTL models (e.g., MMoE, PLE) did not consider feature interaction during the optimization. Through feature importance analysis across various tasks in MTL, we have observed that the same feature can have significantly different importance values across different tasks in MTL, which is called divergence phenomenon in this paper. Therefore, in MTL tasks it is essential to consider feature interactions to learn better representations.

\begin{figure}[H]
  \setlength{\abovecaptionskip}{1pt}
  \includegraphics[width=0.85\linewidth]{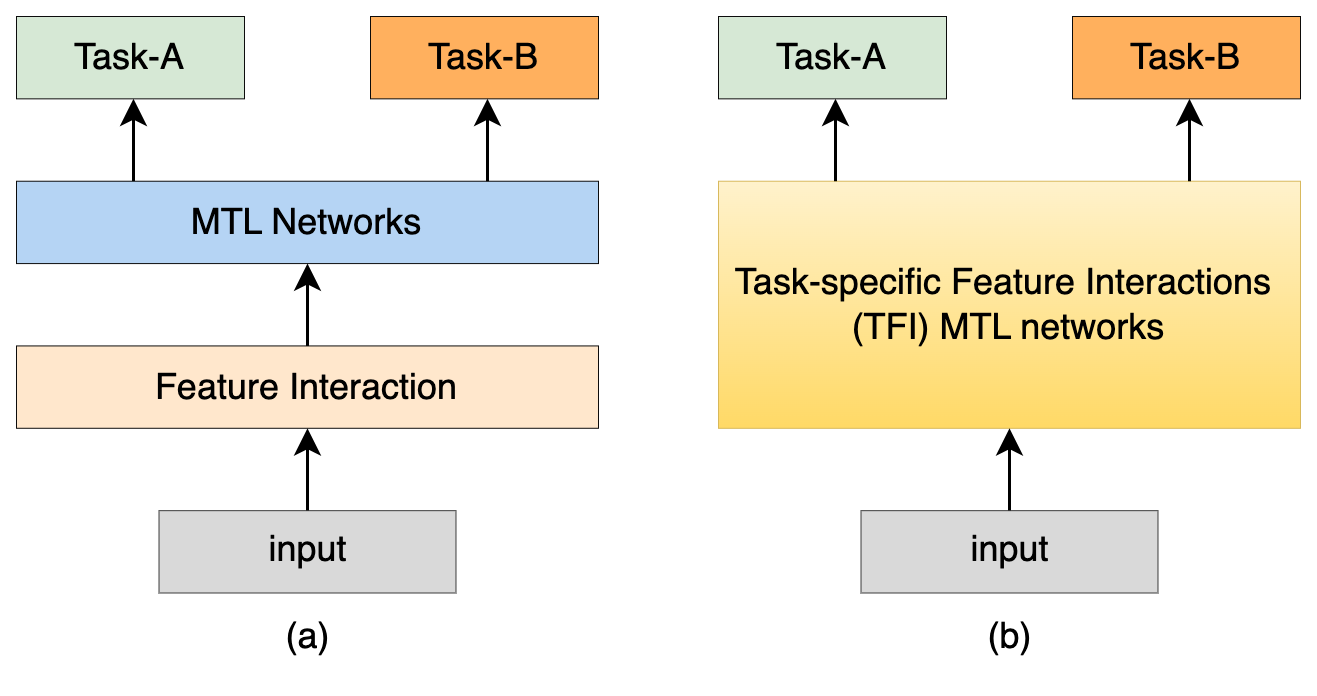}
  \caption{(a) Serial Feature Interaction approach in MTL (SFM). (b) Task-specific Feature Interactions (TFI) MTL model.}
  \label{Fig.mtl}
\end{figure}

A straightforward approach is to perform feature interaction in a serial manner, which means conducting feature interaction based on input embeddings. Subsequently, MTL networks can be applied based on the resulting feature interaction outcomes, which is introduced in Fig.\ref{Fig.mtl} (a). However, this is not the most optimal method for achieving the best performance. It is crucial to design a more powerful and efficient model capable of handling feature interactions in MTL and learn task-specific feature interaction representations for each task in MTL.

To achieve this goal, we propose a novel MTL model named Deep Multiple Task-specific Feature Interactions Network (DTN), which ingeniously considers feature interaction within MTL. Compared to the serial feature interaction approach method in Fig.\ref{Fig.mtl} (a), DTN utilizes feature interaction modules instead of traditional expert modules, enabling improved learning of both task-shared and task-specific feature interaction representations. Additionally, DTN incorporates diversified multiple feature interactions to capture varied feature interaction representations in MTL, thereby maximizing the effectiveness of the model in achieving optimal results.

To evaluate the performance of DTN, we conduct extensive experiments on both real-world industrial recommender systems and major publicly available datasets, including census-income \cite{misc_census-income_(kdd)_117}, AE \cite{peng2020improving} and Ali-CCP \cite{ma2018modeling}. The experiment results demonstrate that DTN outperforms state-of-the-art MTL models across all datasets, and also significantly better than the serial approach of feature interaction in MTL. Besides, significant improvement of online metrics on a large-scale E-commerce recommender system in Shopee demonstrates the advantage of DTN in real-world recommendation applications.

The contributions of our work are summarized as follows:
\begin{enumerate}
 
    \item[$\bullet$] Through feature importance analysis across various tasks in MTL, we have observed an interesting divergence phenomenon that the same feature can have significantly different importance values across different tasks in MTL.

    \item[$\bullet$] A DTN model with a novel feature interaction approach within MTL and a task-sensitive network is proposed to enhance the efficiency of joint representation learning for each individual task. Meanwhile, the MFI module can also be used independently for single-task learning, such as CTR prediction task. Beyond recommendation applications, DTN is versatile and can be applied to a variety of scenarios. 

    \item[$\bullet$] We conducted extensive offline experiments and online to evaluate the effectiveness of DTN using both industrial and public benchmark datasets. The visualization of DTN also shows that different feature interaction methods result in diversified representations.

\end{enumerate}

\section{Related Work}
Efficient multi-task learning models in recommender systems and feature interaction are two research areas related to our work. In this section, we briefly discuss related works in these two areas.

\subsection{Feature Interaction}
 
Many deep learning based ranking models have been proposed in recent years and it is the key factor for most of these neural network based models to effectively model the feature interactions, such as Factorization Machines (FM) \cite{rendle2010factorization}, DeepFM\cite{guo2017deepfm}, xDeepFM\cite{lian2018xdeepfm}, DCN\cite{wang2017deep}, DCN-M\cite{wang2020dcn}, FiBiNET\cite{huang2019fibinet}, MaskNet\cite{wang2021masknet} and Momonet \cite{zhang2023memonet} etc.

 Recently, MaskNet \cite{wang2021masknet} has been proposed to learn specific representation vectors for each feature considering its context. DHEN \cite{zhang2022dhen} leverages the strengths of heterogeneous interaction modules to learn a hierarchy of interactions across different orders. MemoNet \cite{zhang2023memonet} utilizes a multi-Hash Codebook Network (HCNet) as a memory mechanism to efficiently learn and memorize representations of cross features in CTR tasks. GDCN \cite{wang2023towards} employs a GCN with an information gate to dynamically filter next-order cross features, enhancing performance and stability in capturing high-order interactions. 

However, these feature interaction methods are primarily applied in single-task prediction scenarios, such as CTR prediction, rather than in multi-tasks prediction scenarios. Moreover, they have not considered combining diversified multiple feature interaction methods to achieve optimal results, which is an optimization we address in this paper.

\subsection{Multi-Task Learning in Recommender Systems}

To better leverage various user behaviors, multi-task learning has been widely applied in recommender systems, leading to significant improvements, such as ESMM \cite{ma2018entire}, MMoE \cite{ma2018modeling} , Customized Gate Control (CGC) \cite{tang2020progressive} and PLE \cite{tang2020progressive}.

The ESSM \cite{ma2018entire} introduces CTR, CVR task and shares embedding parameters to enhance the accuracy of CVR task. MMoE \cite{ma2018modeling} integrates shared experts using different gating networks for each specific task. PLE \cite{tang2020progressive} explicitly separates task-shared and task-specific experts and employs a progressive routing mechanism to gradually extract and segregate deeper semantic knowledge. 
 
Compared to MMoE and PLE, which did not consider feature interactions during optimization, the DTN model introduced in this paper incorporates task-specific feature interactions. Furthermore, DTN integrates multiple diversified feature interaction methods, such as MaskNet, GDCN, MemoNet and others, into the MTL networks to enhance the learning representation of diversified feature interactions specific to each task. This leads to significant improvements over PLE in real-world recommender systems.

\section{Divergence Phenomenon of Feature Importance}

In this section, we will introduce the Permutation Feature Importance analysis methods, which can help us understand the importance of each feature for a specific task in ranking models. Moreover, we will discuss the divergence phenomenon, where the importance of the same feature can vary significantly across different tasks in MTL. This observation highlights the significance of considering task-specific feature interactions in MTL.

\subsection{Permutation Feature Importance}

Feature effectiveness plays a crucial role in complex systems, particularly in ranking models used in real-world industrial recommender systems. In these systems, the identification of the most important feature sets is vital as it enables us to select key features that can maximize online prediction performance with limited computational resources. Feature importance can be obtained through various methods, including Dropout \cite{chang2017dropout}, FSCD \cite{ma2021towards}, and Permutation Feature Importance \cite{breiman2001random}. We utilize permutation feature importance to obtain feature importance, which is widely used in real-world ranking models.

Permutation feature importance evaluates the importance of features by randomly permuting their values and measuring the change in model performance, showing how dependent the model's performance on the feature. We employ the Feature Importance (FI) metric, which quantifies the change in the Area Under the Receiver Operating Characteristic Curve (AUC) when the values of a feature are permuted. The FI of feature $X_i$ can be described as follows:
\begin{equation}
  FI(X_i) = AUC_{base} - AUC_{p}(X_i),
\end{equation}
where $AUC_{base}$ is the AUC of base model, and $AUC_{p}(X_i)$ represents the AUC of model when the feature $X_i$ is permuted on the test data. The higher the importance value $FI(X_i)$, the more critical that feature is to the model. We use this method to obtain the importance of features for multi-tasks in MTL.

\subsection{Divergence Phenomenon of Feature Importance in MTL}

There are multiple different tasks in MTL as we introduced. Hence, we can independently permute the feature $X_i$ in the test data for each Task $k$ to obtain the importance of $X_i$ in MTL across various tasks. The feature importance of $X_i$ for Task $k$ can be described as follows:
\begin{equation}
  FI^k(X_i) = AUC^k_{base} - AUC^k_{p}(X_i),
\end{equation}
where $AUC^k_{base}$ is the baseline AUC of task $k$ when no feature is permuted on the test data, and $AUC^k_{p}(X_i)$ represents the AUC of task $k$ when the feature $X_i$ is permuted on the test data. The higher the importance value $FI^k(X_i)$, the more critical that feature is to the task $k$.

There are nearly 400 features in the ranking model of Shopee's E-commerce Recommender System. We analyzed the importance of all these features for both click-through rate (CTR) and conversion rate (CVR) tasks in our multi-task learning (MTL) ranking model using Permutation Feature Importance. The results are shown in Table \ref{tab:auc_uplift}, where we used $FI^k(X_i)$ and $Rank^k_{FI}(X_i)$ to assess the importance of feature $X_i$ for task $k$. The ranking, denoted by $Rank^k_{FI}(X_i)$, represents the descending order of the features based on their corresponding $FI^k(X_i)$ values for each task k. We presented several examples to demonstrate the divergence phenomenon of feature importance in Multi-Task Learning (MTL) in Table \ref{tab:auc_uplift}.

\begin{table}[h]
  \setlength{\abovecaptionskip}{1pt}
\centering
\caption{Feature Importance (FI) of features for task CTR/CVR}
\begin{tabular}{lcccc}
\toprule
feature  &  CTR FI & $Rank^{CTR}_{FI}$ & CVR FI & $Rank^{CVR}_{FI}$ \\
\midrule 
item\_price   &       0.0076  &       2       &       0.0077  &       1       \\
image\_ebd     &       0.0070   &       3       &       0.0001 &       334     \\
item\_ctr\_48h    &       0.0054  &       9       &       0.0003  &       216     \\
realtime\_discount       &       0.0045  &       23      &       0.0070   &       3   \\
atc\_items\_50        &       0.0001 &       352     &       0.0046  &       19      \\
order\_cnt\_24h    &       0.0001 &       367     &       0.0058  &       11      \\
atc\_cnt\_12h     &       0.0001 &       386     &       0.0024  &       83      \\
\bottomrule
\end{tabular}
\label{tab:auc_uplift}
\end{table}

We can observe that the features $item\_price$ and $realtime\_discount$ (real-time discount ratio of each item) exhibit high importance in both the CTR and CVR tasks. However, the feature $item\_ctr\_48h$ (ctr of the item in recent 48 hours) and $image\_ebd$ (pre-trained embedding of item cover) show high importance in CTR but low importance in CVR. Conversely, the feature $order\_cnt\_24h$ (user order count in the recent 24 hours) and $atc\_items\_50$ demonstrate high importance in CVR but low importance in CTR. Certainly, there are many other features that can demonstrate this divergence phenomenon of feature importance across different tasks in MTL.

\begin{figure}[H]
  \setlength{\abovecaptionskip}{1pt}
  \includegraphics[width=0.95\linewidth]{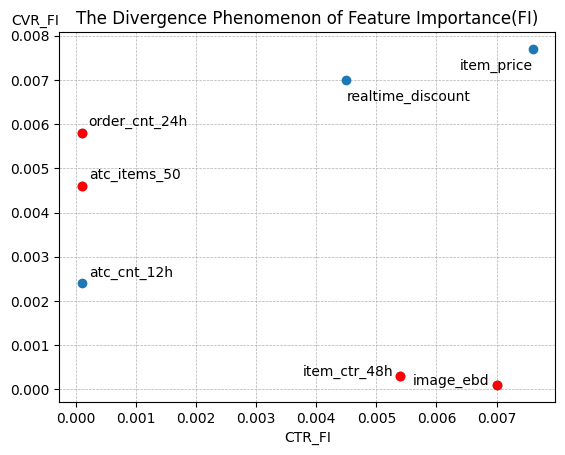}
  \caption{The Divergence Phenomenon: same feature have significantly different feature importance across CTR, CVR task in MTL, such as $order\_cnt\_24h$ and $item\_ctr\_48h$.}
  \label{Fig.feature}
\end{figure}

This phenomenon can also be observed intuitively through Fig. \ref{Fig.feature}, where the $x$-axis represents the feature importance of $X_i$ in the CTR task, and the $y$-axis represents the feature importance of $X_i$ in the CVR task. Through the positions of feature $item\_ctr\_48h$, $image\_ebd$, $order\_cnt\_24h$ and $atc\_items\_50$, directly shows the divergence of feature importance.

This phenomenon illustrates the divergence of feature importance, wherein the significance of the same feature can vary substantially across different tasks in MTL. This divergence extends to feature interactions, which are constructed by combining basic features to generate high-order representations. Their effectiveness and importance also vary across tasks in MTL. Therefore, introducing task-specific feature interaction mechanisms is essential, as it enables the learning of task-specific interaction representations, thereby improving model performance and generalization. In Section \ref{subsec:vdtn}, we also offer a comprehensive analysis of the variations in result representations among different feature interaction methods across multiple tasks in MTL.

\section{Modeling Approaches}

In this section, we present our solution to address the divergence phenomenon called the Deep Multiple Task-specific Feature Interactions Network (DTN) model, which incorporates a novel structure design. Firstly, we introduce the basic serial approach for handling feature interaction in the context of MTL. Secondly, we introduce the Task-Specific Feature Interaction (TFI) model, which encompasses both task-shared and task-specific feature interactions. Lastly, we provide a detailed description of the proposed DTN structure.

\subsection{SFM: Serial Feature Interaction Approach in MTL}

As previously mentioned, the architecture of MTL models in RS, such as Customized Gate Control (CGC) model introduced in PLE and MMoE, directly build multi-expert and gating-based MTL networks based on input embeddings without consider feature interaction during the optimization process.

As introduced in Fig. \ref{Fig.mtl} (a), the Serial Feature Interaction in MTL (SFM) is a serial approach to directly incorporate feature interaction based on input embeddings, then building the MTL networks based on output of feature interaction module. But SFM neglects to consider task-specific feature interaction outcomes for each specific task.

\subsection{TFI: Task-Specific Feature Interaction}

As shown in Fig. \ref{Fig.mtl} (b), based on PLE model we propose TFI model, which integrates feature interaction into MTL networks. 

By learning task-specific feature interaction representation in MTL networks, it facilitates the direct utilization of personalized learning via Feature Interaction Modules (FIMs) to capture the unique Feature Interaction patterns of each Task. Specifically, we have the flexibility to choose any feature interaction method that we want to incorporate, such as DCN, MaskNet, GDCN, MomoNet, etc.

Through the proposed FIMs module, TFI model can learn task-specific feature interaction representations, leading to a more effective expression of feature interactions.

 \begin{figure*}
   \setlength{\abovecaptionskip}{1pt}
   \includegraphics[width=1.0\linewidth]{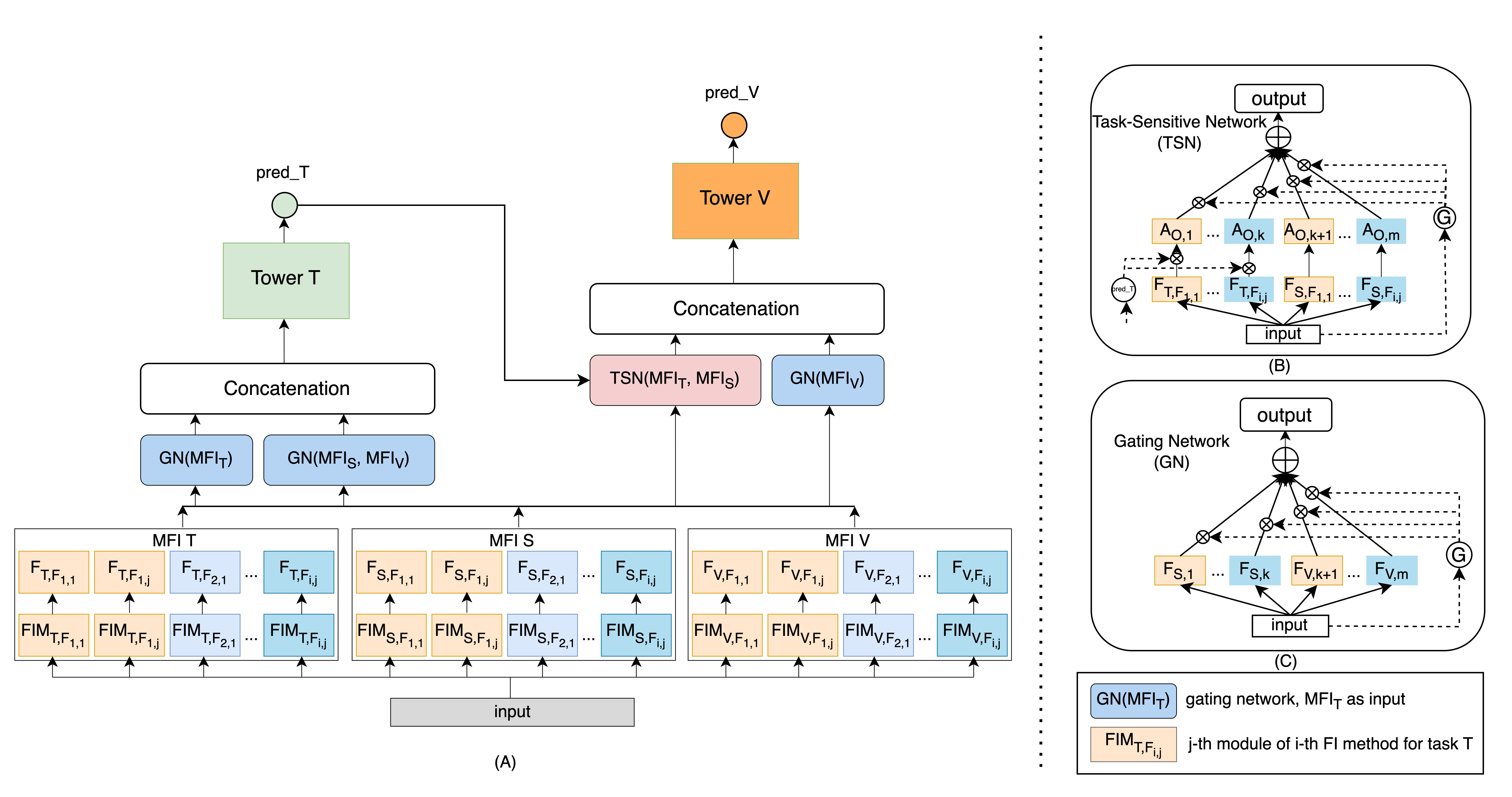}
   \caption{\  (A)\ Architecture overview of DTN, which utilizes Multiple Feature Interaction (MFI) module and Task-Sensitive Network (TSN) to learn task-specific feature interaction representations. \ (B)\  TSN takes into account the sequential nature of tasks and leverages the representations of preceding tasks more effectively. \ (C)\  Gating Network (GN) is employed to compute the weighted sum of feature interaction representations. }
   \label{Fig-dtn}
 \end{figure*}

\subsection{DTN: Deep Multiple Task-specific Feature Interactions Network}
 
To better address the variations in feature representations resulting from different feature interaction methods, we introduce the Deep Multiple Task-specific Feature Interactions Network (as shown in Fig. \ref{Fig-dtn}), which includes Multiple diversified Feature Interactions (MFI) modules and Task-Sensitive Network (TSN).

\subsubsection{Multiple diversified Feature Interactions }
To enhance the model's ability of learning feature interaction representations, we using Multiple diversified Feature Interactions (MFI) modules to learn diversified feature interaction representations based on diversified feature interaction methods. As illustrated in Fig. \ref{Fig-dtn}, the diagram depicts task-specific and task-shared MFI modules based on input layer. 

For each MFI module, we have the flexibility to choose multiple feature interaction methods that we want to incorporate, such as SENet, DCN-M, MaskNet, MemoNet, GDCN, and any other newly emerged feature interaction methods. 

The MFI module for task CVR can be expressed as follows:
\begin{equation}
  M_{V}(x) = [F_{V,F_{1,1}},...,F_{V,F_{1,j}},...,F_{V,F_{i,j}}],
\end{equation}
where $x$ is the input representation, and $F_{V,F_{i,j}}$ represents output vector of the $j$-th module of the $i$-th feature interaction method for task CVR. And the MFI module for the CTR task can be expressed as $M_{T}(x)$, while the shared MFI module is denoted as $M_{S}(x)$.

\subsubsection{Task-Sensitive Network } In order to select different feature interaction approaches for each tasks, we use the gating network to learn the precise weight of each feature interaction approach for each task. The gating network (GN) is based on a single-layer feed-forward network with softmax as the activation function, input as the selector to calculate the weighted sum of the selected vectors, i.e, the output vector of feature interaction modules.

Moreover, it is important to acknowledge the sequential nature of multiple tasks in MTL, such as the requirement for users to click on an item before being able to place an order, as well as the need for users to click on a video in video recommendation scenarios in order to have the duration of play.

Therefore, for the subsequent task $V$, we introduce a more explicit signal by multiplying the prediction score of the preceding task $T$ with the output of $MFI_T$ to obtain the final representation of the preceding task $T$. This allows us to consider the interdependencies between tasks and enhance the modeling of subsequent tasks. We refer to this approach as Task-Sensitive Network (TSN).

Specifically, TSN for task $V$ contains two parts of input: the predicted score of $T$ task $pred\_T$, and output vecotr of MFI modules $ M = [M_{T}(x), M_{S}(x), M_{V}(x)]$. By multiplying the $pred\_T$ with the relevant MFI module, we obtain $A_O = [pred\_T\cdot M_{T}(x), M_{S}(x)]$, with each vector in $A_O$ denoted as $A_O^i$. For the $T$ task, only the gating network (GN) is utilized since it doesn't involve any pre-task, unlike the $V$ task.

In order to effectively retain the task-specific information learned for each task, our approach utilizes two gates for each task: one GN for the task-specific MFI and another for the task-shared and other task's MFI components. For task $V$, the task-shared and other task's MFI components we calculate the output $S_O$ as follows:

\begin{equation}
S_O = \sum_i^m g_O^i A_O^i
\end{equation}
where $g_O^i$ represents the weights of FIMs learned by the gating network, and $m$ is the total number of $A_O^i$. Similarly, we have the output $S_V$ for task-specific MFI components as:
\begin{equation}
S_V = \sum_i^n g_V^i M_{V}(x)
\end{equation}
where $n$ is the total number of FIM modules in $M_{V}(x)$, so we have the output $C_V$ for task $V$ as: $C_V = Concat[S_O, S_V]$. 

And the tower network is a multi-layer network with width and depth as hyper-parameters. The prediction of task $V$ is:
\begin{equation}
  y_V = t_VC_V,
\end{equation}

\noindent where $t_V$ denotes the tower unit of task $V$. 

Analogously, we have the prediction of $T$ task is:
\begin{equation}
  y_T = t_TC_T,
\end{equation}
\noindent where $t_T$ denotes the tower unit of task $T$. The only difference is that the task $T$ doesn't have a pre-task, so only the gating network (GN) is used for it.

Through incorporating multiple diversified feature interaction methods, DTN enhances the model's ability to capture diverse and nuanced feature interactions. This ensemble strategy not only provides a more robust solution but also offers flexibility in adapting to various data characteristics and task requirements.

\begin{table*}
  \setlength{\abovecaptionskip}{1pt}
\centering
\caption{Performance comparison on CTR, Add-To-Cart (ATC) and CVR Task in Shopee dateset. We currently employ three newest feature interaction methods in DTN, such as GDCN, MemoNet and MaskNet for illustration. DTN-trim is the trimmed version of DTN that selects high-importance feature interactions for each task based on gating weights in Shopee dataset, which will introduce in section \ref{subsec:fis} .}

\begin{tabular}{lcccccccccc}
\toprule
\multirow{2}{*}{Models}  & Feature Interactions  & \multicolumn{3}{c}{CTR} & \multicolumn{3}{c}{ATC} & \multicolumn{3}{c}{CVR}  \\ 
& & AUC & RelaImpr & LogLoss & AUC & RelaImpr & LogLoss& AUC & RelaImpr & LogLoss\\
\midrule
Shared-Bottom & - & 0.7341	&	-	&	0.5631	&	0.7759	&	-	&	0.4931	&	0.8029	&	-	&	0.4627 \\
\hline
MMoE & - & 0.7363	&	+0.93\%	&	0.5571	&	0.7786	&	+0.97\%	&	0.4881	&	0.8041	&	+0.39\%	&	0.4572 \\
\hline
PLE & - & 0.7401	&	+2.56\%	&	0.5513	&	0.7800	&	+1.48\%	&	0.4834	&	0.8052	&	+0.75\%	&	0.4516 \\
\hline

 &  GDCN & 0.7412	&	+3.03\%	&	0.5473	&	0.7801	&	+1.52\%	&	0.4830	&	0.8067	&	+1.25\%	&	0.4462 \\
SFM & MemoNet & 0.7424	&	+3.54\%	&	0.5429	&	0.7808	&	+1.77\%	&	0.4805	&	0.8083	&	+1.78\%	&	0.4404 \\
 &  MaskNet & 0.7448	&	+4.57\%	&	0.5341	&	0.7816	&	+2.06\%	&	0.4777	&	0.8095	&	+2.17\%	&	0.4361 \\
\hline
 &  GDCN &  0.7424	&	+3.54\%	&	0.5429	&	0.7814	&	+1.99\%	&	0.4784	&	0.8078	&	+1.61\%	&	0.4422\\
TFI & MemoNet & 0.7440	&	+4.22\%	&	0.5371	&	0.7813	&	+1.95\%	&	0.4788	&	0.8076	&	+1.55\%	&	0.4429 \\
 &  MaskNet & 0.7473	&	+5.63\%	&	0.5249	&	0.7844	&	+3.08\%	&	0.4678	&	0.8107	&	+1.38\%	&	0.4318\\
\hline
DTN(w/o TSN) & MFI & 0.7481	&	+5.98\%	&	0.5220	&	0.7845	&	+3.11\%	&	0.4674	&	0.8143	&	+3.92\%	&	0.4188 \\
\frenchspacing{DTN}\ & MFI & \textbf{0.7483}	&	\textbf{+6.06\%}	&	\textbf{0.5219}	&	\textbf{0.7850}	&	\textbf{+3.29\%}	&	\textbf{0.4637}	&	\textbf{0.8153}	&	\textbf{+4.09\%}	&	\textbf{0.4160} \\
DTN-trim & MFI+trim & 0.7482	& +6.02\% & 	0.5231 & 0.7847	& +3.19\%	& 0.4667	& 0.8147 & +3.89\% & 0.4174\\
\bottomrule
\end{tabular}
\label{tab:offline_result}
\end{table*}

\section{Experiment}

In this section, we conduct extensive offline and online experiments on both Shopee's large-scale recommender system and public benchmark datasets. These experiments aim to evaluate the effectiveness of the proposed models. The results demonstrate that our approach outperforms state-of-the-art MTL methods.

\subsection{Evaluation on the E-commerce Recommender System in Shopee}

In this subsection, we perform offline and online experiments on various tasks within Shopee's E-commerce recommender system to assess the performance of the proposed models.

\subsubsection{Dataset} We collected traffic logs from the online E-commerce recommender system in Shopee, of which one week’s samples are used for training and samples of the following day for testing. The size of training and testing set is about 6.3 billions and 0.9 billion respectively. We focus on three tasks in this dataset, namely CTR, ATC, and CVR, which model user preferences.





\subsubsection{Models for Comparisons} In the experiment, we compare SFM, TFI and DTN model with the Shared-Bottom, MMoE and PLE model. Considering that DHEN is a serial method which will greatly increase model latency, DTN does not use this feature interaction method. In our DTN model we employed the TSN for the ATC target as well for better performance. We introduce three newest feature interaction methods used in DTN as follows:

\begin{enumerate} 
  \item[$\bullet$] \textbf{GDCN: }  GDCN \cite{wang2023towards} utilizes low-rank methods to approximate feature interactions in subspaces, achieving a better trade-off between model performance and latency. We choose GDCN as a feature interaction module instead of DCN \cite{wang2017deep} and DCN-M\cite{wang2020dcn}.
  
  \item[$\bullet$] \textbf{MemoNet: }  MemoNet \cite{zhang2023memonet} use multi-hash codebooks to memorize and learn the feature interaction. In order to keep the same params as other feture interaction methods, we only choose part of user features and item features with top feature importances to cross in MemoNet, which may not result in the optimal performance for this feature interaction method.
  
  \item[$\bullet$] \textbf{MaskNet: } MaskNet \cite{wang2021masknet} has been proposed to learn specific representation vectors for each feature considering its context, which we think should be very useful in ranking models in RS.
  
\end{enumerate}

\subsubsection{Offline Experiment Setup} \label{subsubsec:implement} In our experiment, CTR, ATC and CVR prediction task are binary classification tasks, trained by cross-entropy loss and evaluation metric by AUC. Each FIM outputs dimensions in DTN is 512, and task-sensitive network. The output dimension of each feature interaction module in SFM, TFI, and DTN remains consistent at 512.  We use Adam\cite{kingma2014adam} optimizer in our experiment, and the default learning rate is set to 1e-3. The batch size is set to 2048. The embedding dimentions is 16. Unless otherwise specified, all activation functions are ReLU. We use 2-layer gated cross layers for our GDCN interaction. And use part of user and item features for our MemoNet method, with 2 HCNets, and we apply 2-order interactions inside the MemoNet. For MaskNet method, the mask gate block we use a hidden layer dimensions of 640, apply layer normalization to it, and then compress it to the same dimension of 512 at output. We use 12 TFI for each model and control the parameter size of each TFI to be similar. The overall parameter is around 1.6E8 for different models to make a fair comparison, except for $DTN\_trim$ model, which reduced about 9\% of parameters than DTN. 

\subsubsection{Evaluation Metrics} In our experiments, we use AUC (Area Under ROC) and LogLoss as the evaluation metric. AUC has an upper bound of 1, and a higher value indicates better performance. RelaImp, as described in work \cite{xie2021deep}, is used as another evaluation metric to measure the relative AUC improvements over the corresponding baseline model. Since AUC is 0.5 from a random strategy, we can remove the constant part of the AUC score and formalize the RelaImp as:
\begin{equation}
  RelaImp = (\frac{AUC(Measured\ Model) - 0.5}{AUC(Base\ Model) - 0.5} - 1) \times 100\%,
\end{equation}

\begin{table*}
    \setlength{\abovecaptionskip}{1pt}
\centering
\small{
\caption{Experiment Results on Public Datasets}
\begin{tabular}{l|cccccccccccc}
\toprule
\multirow{2}{*}{\textbf{Models}}  & \multicolumn{2}{c}{\textbf{Census Task1}} & \multicolumn{2}{c}{\textbf{Census Task2}}  & \multicolumn{2}{c}{\textbf{AE CTR}} & \multicolumn{2}{c}{\textbf{AE CVR}} & \multicolumn{2}{c}{\textbf{Ali-CCP CTR}} & \multicolumn{2}{c}{\textbf{Ali-CCP CVR}} \\ 
& AUC & RelaImpr   & AUC & RelaImpr  & AUC & RelaImpr & AUC & RelaImpr   & AUC & RelaImpr  & AUC & RelaImpr \\
\midrule
Shared-Bottom & 0.9361 & - & 0.9915 & - & 0.7173 & - & 0.8111 & - & 0.6061 & - & 0.6121 & - \\
MMoE & 0.9410 & +1.12\% & 0.9926 & +0.22\% & 0.7216 & +1.98\% & 0.8126 & +0.48\% & 0.6094 & +3.1\% & 0.6162 & +3.6\% \\
PLE & 0.9521	& +3.6\%  & 	0.9945  &	+0.61\%  &  0.7211 & +1.75\%	& 0.8144	& +1.06\% &	0.6112 &	+4.8\%	 & 0.6183	& +5.5\% \\
DTN & \textbf{0.9560}	& \textbf{+4.56\%} & \textbf{0.9952} & \textbf{+0.75\%}	& \textbf{0.7243} &	\textbf{+3.22\%} &	\textbf{0.8207} & \textbf{+3.09\%}	& \textbf{0.6153}	& \textbf{+8.6\%}	& \textbf{0.6231}	& \textbf{+9.8\%} \\
\bottomrule
\end{tabular}
\label{tab:overalperformance}
}
\end{table*}

\subsubsection{Evaluation on Shopee dataset} 

The experiment results from Shopee's offline evaluations are summarized in Table \ref{tab:offline_result}. From Table \ref{tab:offline_result} we can conclude that applying feature interactions such as GDCN, MemoNet and MaskNet in SFM has demonstrated an uplift in AUC metrics. Further enhancements were achieved by applying the task-specific feature interaction module in TFI model, which allowed us to learn the task-specific feature interaction representations. The most significant improvements were observed with our DTN model, in which we integrate three feature interaction methods in this experiment, resulting in the highest metric uplifts for CTR, ATC, and CVR prediction tasks. And we also include a DTN-trim version for comparison, where we utilize gating weights to trim the DTN. This shows the effectiveness of MFI module in DTN, where diversified task-specific feature interaction modules are selectively utilized to optimize performance across varied tasks.

\subsubsection{Online A/B Testing} 
The DTN model has been applied within Shopee's E-commerce recommender system and rigorously tested to evaluate its effectiveness in a live environment. We conducted a 7-days online A/B testing to directly compare the performance of DTN against the existing online model, which is PLE. The results of this testing are summarized in Table \ref{tab:online_result}, which details the improvements achieved by our model. The results indicate a significant uplift in clicks by +3.28\%, and orders + 3.10\% and GMV (Gross Merchandise Volume) +2.70\%. The result shows the improvements of DTN model compared to state-of-the-art MTL models.

\begin{table}[H]
  \setlength{\abovecaptionskip}{1pt}
\centering
\caption{Improvement over PLE on Online A/B Test.}
\begin{tabular}{lccccc}
\toprule
Model  & \ Clicks  & \ Orders & GMV \\
\midrule
DTN & \textbf{+3.28\%}  & \textbf{+3.10\%}  & \textbf{+2.70\%} \\
\bottomrule
\end{tabular}
\label{tab:online_result}
\end{table}

\subsection{Evaluation on Public Datasets} 
In this subsection, we conduct experiments on three public benchmark datasets to evaluate the effectiveness of DTN in scenarios beyond recommendation.

\subsubsection{Datasets}

\begin{enumerate}

  \item[$\bullet$] \textbf{Census-Income Dataset\cite{misc_census-income_(kdd)_117}} is a public dataset extracted from the 1994 census brureau database, which provide 40 features for training and two tasks for prediction, one task is whether the income exceeds 50K, the other task is to predict whether the person is married.
  
  \item[$\bullet$] \textbf{Public AE Dataset\cite{peng2020improving}} is a public dataset which contains 22,326,719 training samples and 9,342,708 test samples gathered from real-world traffic logs of AliExpress. We will predict the CTR and CVR task inside this dataset.
  
  \item[$\bullet$] \textbf{Ali-CCP Dataset\cite{ma2018modeling}} is a public dataset from Taobao's traffic logs with 42,300,135 training samples and 43,016,840 test samples included. It also focuses on CTR and CVR predictions.

\end{enumerate}

\subsubsection{Implementation Details} 
 
The setup for census-income dataset is the same as \cite{ma2018modeling}. For AE and Ali-CCP dataset, we also compare DTN with Shared-Bottom, MMoE and PLE models, which is same as \cite{tang2020progressive}. Same as implement details introduced in Offline Experiment Setup section, we implement all the models with Tensorflow in our experiments and control the overall parameter size of each model to have a fair comparison.

\subsubsection{Experiment Results} Experiment results on public datasets are shown in Table \ref{tab:overalperformance}. We can observe that the DTN model outperforms PLE on the three datasets. In this experiment, we employed the GDCN, MemoNet, and MaskNet these three feature interaction methods in the DTN model. This version of DTN have already surpassed the state-of-the-art PLE model in performance. However, the DTN model still has the potential to incorporate more additional effective feature interaction methods for further improvement.

Combining previous experiments on the industrial dataset and online A/B tests conducted within Shopee, the DTN model has demonstrated extraordinary performance in learning representations through multiple diversified feature interaction methods, resulting in the most significant performance improvement.

\subsection{Feature Interaction Selection}
\label{subsec:fis}
We know that different task-specific feature interaction methods indicate different feature representations, and we utilize the feature interaction selection for each task in order to get the task-specific feature representation more efficiently. We retrieve the gating weights of our feature interaction methods of task-specific MFI module and other MFI modules for each task. 

\begin{figure}[h]
  \setlength{\abovecaptionskip}{1pt}
  \centering 
  \includegraphics[width=0.95\linewidth]{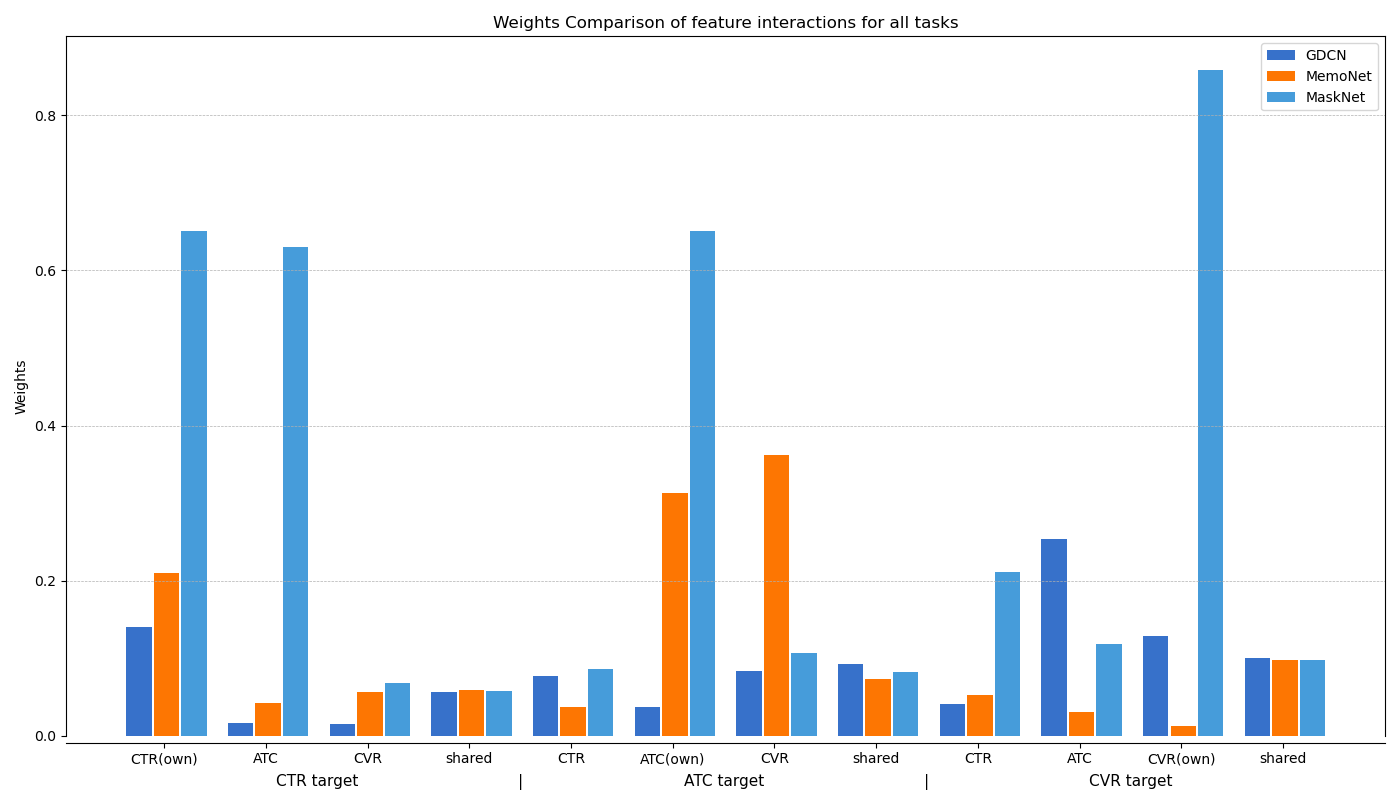}
  \caption{Visualization of FIM Weights for CTR, ATC, and CVR Tasks in DTN.}
  \label{figure-fi-selection}
\end{figure}

As illustrated in \ref{figure-fi-selection}, we compare the contributions of MaskNet, MemoNet and GDCN in DTN that trained in Shopee data. We can observe that for each tasks, their respective MaskNet feature interaction modules contribute the most. And for the ATC task, MemoNet demonstrates the best performance in its own MFI module as well as in the CVR MFI module. For the CVR task, GDCN outperforms MemoNet in its own MFI module.

In order to consider the online serving efficiency of DTN, we can cut out some less important feature interaction modules based on gating weights, such as GDCN in task-specific MFI module of CVR task. This approach allows different tasks to utilize the most suitable feature interaction modules, also helps reduce resource consumption while maintaining an acceptable level of performance degradation.

\begin{figure}[h]
  \setlength{\abovecaptionskip}{1pt}
  \centering 
  \includegraphics[width=0.95\linewidth]{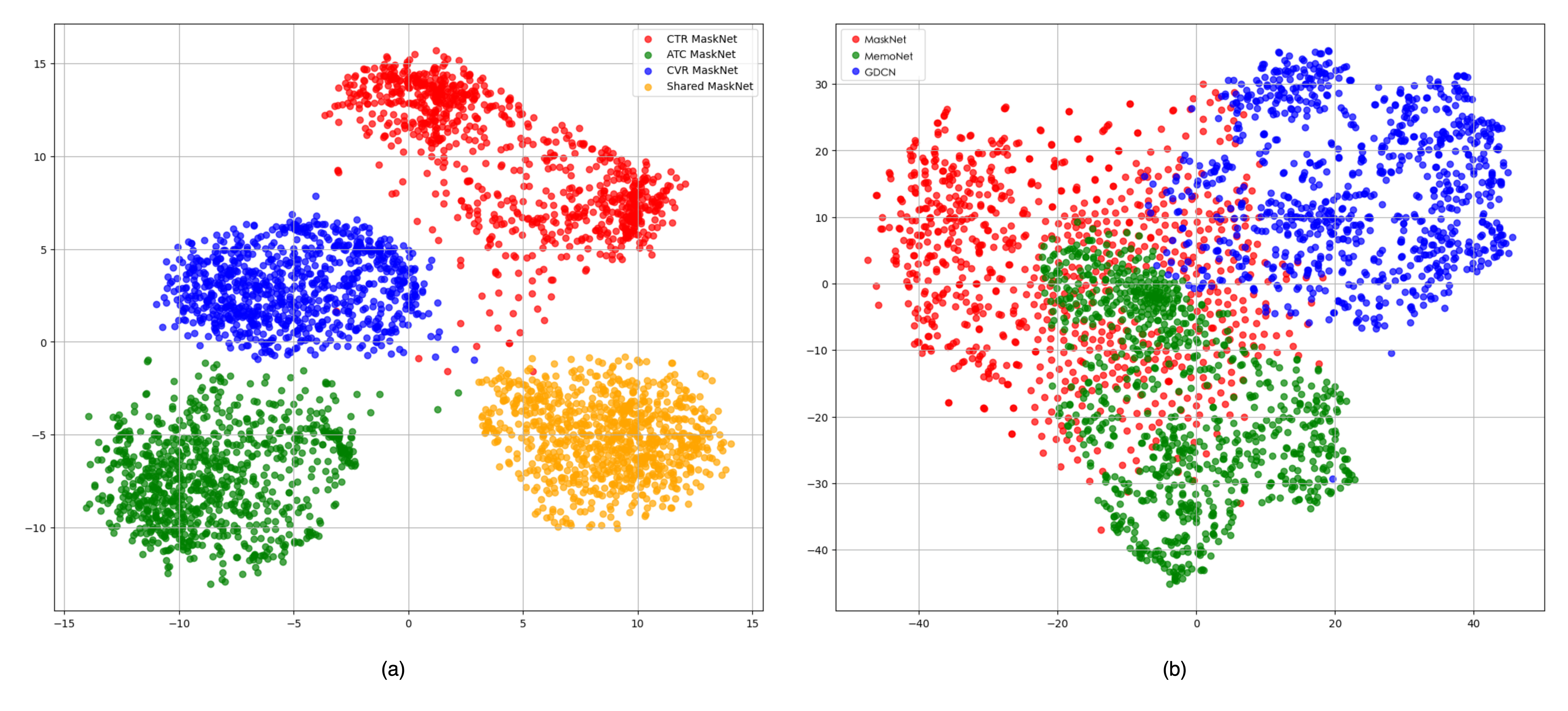}
  \caption{ Visualization of DTN: (a) DTN can learn task-specific representations with same FIM method. (b) Different FIM learn diversified representatioins for same task. }
  \label{fig-dtn-fi}
\end{figure}

\subsection{Visualization of DTN}
\label{subsec:vdtn}
Finally, we visualize the task-specific feature interaction representations of the feature interaction modules in DTN on Shopee dataset. It demonstrates that DTN can learn diversified task-specific feature interaction representations, which enhances the model's performance.

We start by visualizing the task-specific feature interaction representations from the feature interaction modules across the CTR, ATC, and CVR tasks on Shopee dataset. We randomly select 1,000 samples from the dataset, and taking the MaskNet feature interaction module as example, which have a better performance than MemoNet and GDCN as we showed in Table \ref{tab:offline_result}. Fig. \ref{fig-dtn-fi}(a) illustrates the visualization of the representation of the MaskNet feature interaction method using t-SNE\cite{van2008visualizing} learned by DTN, in which points with same color correspond to the same task. As expected, Fig. \ref{fig-dtn-fi}(a) demonstrates that the representations vectors of task-specific feature interactions exhibit significant differences in comparison to both other tasks and the shared feature interactions. This emphasizes the importance of incorporating task-specific feature interactions.

We then visualize the representations vectors of different feature interaction methods. Taking the CTR task as an example, Fig. \ref{fig-dtn-fi}(b) shows different feature interaction method in DTN can learn diversified feature interaction representations.

\section{Conclusion}
In this paper, we propose a novel MTL model called DTN. Our model incorporates diversified feature interaction methods and task-sensitive network into MTL networks, which results in improved task-specific feature interaction representations and more efficient joint representation learning. We conducted both offline and online experiments using industrial datasets and public benchmarks, and the results consistently showed significant improvements of DTN compared to state-of-the-art MTL models. Moving forward, our future work will focus on exploring the synergy between MTL and feature interactions, aiming to further enhance the efficiency of joint representation learning in MTL.

\bibliographystyle{ACM-Reference-Format}
\bibliography{dtn}










\end{document}